\documentclass{aastex6}
\PassOptionsToPackage{linktocpage}{hyperref}
\usepackage{amsmath}
\def\bfB{\mathbf{B}}
\def\bfE{\mathbf{E}}
\def\bfJ{\mathbf{J}}
\def\bfQ{\mathbf{Q}}
\def\bfP{\mathbf{P}}
\def\bfv{\mathbf{v}}
\def\bfx{\mathbf{x}}
\def\bfk{\mathbf{k}}

\def\calU{\mathcal U}
\def\calH{\mathcal H}

\begin{document}


\title{Alphonons as the quanta of the Alfv\'{e}n waves}



\author{Zohreh Mohammadi \altaffilmark{1} }

\author{Hossein Safari \altaffilmark{2} }

\author{Farhad Zamani\altaffilmark{1}}
  \affil{Department of Physics, University of Kashan, P. O. Box 87317-53153, Kashan, Iran;   \\
	\and
	Department of Physics, University of Zanjan, P. O. Box 45195-313, Zanjan, Iran.}

\and





\begin{abstract}
	
	In a magnetized plasma, both the transverse and
	longitudinal
	Alfv\'{e}n waves carry an average energy density that, for each mode
	of oscillation, resembles the classical energy of a harmonic
	oscillator with unit mass and angular frequency that depends on the
	background magnetic field and plasma density. We employ this fact to
	introduce a pair of canonical variables for the underlying classical
	system, and then use the canonical quantization procedure to express
	the Alfv\'{e}n wave velocity (or the corresponding plasma
	displacement vector) in terms of creation and annihilation
	operators. We thus introduce the concept of ``{\it Alphonons}" as
	the quanta for the Alfv\'{e}n wave packets and study their
	characteristic features. The corresponding quanta (Alphonons) carry
	spin one for the transverse Alfv\'{e}n wave and spin zero for the
	longitudinal Alfv\'{e}n wave.  We obtained behavior of thermodynamic quantities for
	an Alphonon system (e.g., free energy, entropy, specific heat). We showed that for
	the Alphonon system the spectral energy density obeys the Planck's
	black body radiation law.

\end{abstract}




\section{Introduction} \label{sec:intro}

Alfv\'{e}n suggested there must be the Alfv\'{e}n wave extracted
from Magnetohydrodynamics (MHD) equations observable in a
plasma~\citep{Alfven}. Tension within magnetic field lines is the
restoring force of the Alfv\'{e}n wave inside a magnetized plasma.
Lundquist detected the Alfv\'{e}n wave~\citep{Lundquist}, and Bostick
and Levine used a laboratory setup to study the behavior of this wave~\citep{Bostick}.
Jephcott produced the Alfv\'{e}n wave in a
discharged gas~\citep{Jephcott}. \cite{parker},
\cite{hollweg}, and ~\cite{Roberts}
proposed the propagation of the Alfv\'{e}n wave in space plasma such
as intergalactic, interplanetary medium, and solar plasma,
respectively. \cite{Aschwanden} and \cite{Berthold}
reported the oscillations with millihertz frequencies of the
MHD waves in the solar coronal loops by  analyzing the TRACE
(Transition Region and Coronal Explorer) extreme-ultraviolet
observations. Several attempts have
been done in the field of observations and modeling of the
Alfv\'{e}n and MHD waves in the solar atmospheric plasma (see
e.g., \citep{Safari1,Safari2,Safari3,Safari4,Safari5,farahani2017torsional,ajabshir,Erdelyi,Gruszecki,Kaghashvili,Jess,Srivastava,Pascoe,Andries,Arregui,Moortel}).

Most of the physical characteristics of the gaseous plasma can be
applied in the solid-state plasma. The primary difference is the
number density which is typically large in a solid-state plasma; for
example, it is typically $10^{22}\,{\rm cm}^{-3}$ in metals,
$10^{16}\,{\rm cm}^{-3}$ in semiconductors, $10^{12}\,{\rm
	cm}^{-3}$ in a laboratory plasma, and $10^9\,{\rm cm}^{-3}$ in
solar coronal plasma (see e.g.,~\citep{Bowers,Williams}).
Propagation of the Alfv\'{e}n wave in the solid-state plasma is
detected  for example in bismuth~\citep{Williams,Kirsch},
in graphite~\citep{Nakamura}, and in metals~\citep{Fisher}.

Quasiparticles and collective excitations are introduced when the particles in solid or fluid show different treatment with those in
free space with weak interactions. Several quasiparticles and
collective excitations have been introduced in the physical systems
to simplify the complicated interactions between the particles.
Chargeon~\citep{charjeon}, configuron~\citep{Configuron}, electron quasiparticle~\citep{Electron},
exciton~\citep{exciton1,exciton2},
orbiton~\citep{Orbiton}, phason~\citep{Phason}, phonon, plasmaron~\citep{Plasmaron1,Plasmaron2}, spinon~\citep{Spinon2}, wrinklon~\citep{Wrinklon}, etc, are a list of the
quasiparticles and collective excitations were mostly used in physics.

Song and Lysak defined Alfv\'{e}nons as the macro (quasi) particle,
to give particle like descriptions of both transverse (shear) and
longitudinal (compressional) Alfv\'{e}n waves~\citep{song}. They
explained the driven reconnections in the solar winds as the
interaction of the Alfv\'{e}nons and embedded magnetic field.

Phonons are the quantum of the vibrational waves (sound) or their
equivalent normal modes in a lattice. When a lattice vibrates, atoms
or molecules oscillate around their equilibrium locations and these
oscillations are well known as the normal modes of the lattice.
Phonons are bosonic quasiparticles~\citep{Feynman}. In
three-dimensions, each mode associated with three possible
polarizations, two transverse and one longitudinal. We say that
longitudinal (acoustic) and transverse (optical) phonons have spin
$0$ and $1$, respectively~\citep{Levine}.

An important application of phonons in the artificial black holes
(e.g. acoustic black hole) is given by Unruh~\cite{Unruh}. Since
this pioneering work, several attempts have been made to detect the
Hawking radiation emitted at the horizon of a sonic black hole. In a
sonic black hole the phonons (related to the acoustic wave in
super-sonic flow) can not escape the flow at the horizon of the
black hole and Hawking radiation is emitted from the horizon which
is made by phonons~\citep{Garay,Barcelo,Recati,Zapata,Schutzhold,Rousseaux,Weinfurtner,Steinhauer}.

In a similar analysis, Gheibi, Safari, and Innes introduced the
Alfv\'{e}nic black hole based on the magnetohydrodynamics
approach~\citep{gheibi}. They showed that similar to sonic black
holes, which trap phonons and emit Hawking radiation at the sonic
horizon where the flow speed changes from super- to sub-sonic, in
the horizon of Alfv\'{e}nic  black holes, the Alfv\'{e}n waves will
be trapped and emit Hawking radiation made of quantized vibrations
similar to phonons, for which they coined the name ``{\textit Alphonons}".
They also defined the magnephonons as a new quasi-particle for
Hawking radiation emitted from the horizon of the magnetoacoustic   black hole.

Here, we investigate the physical properties of the Alphonons as the
quanta of the Alfv\'{e}n waves. To do this end, we first observe
that each mode of oscillation of the Alfv\'{e}n waves may be
identified with a harmonic oscillator with unit mass and an angular
frequency that depends on background magnetic field and plasma
density. This allows us to use the machinery of canonical
quantization to express the corresponding velocity field (or
equivalently the plasma displacement) in term of creation and
annihilation operators that define the quanta related to the
Alfv\'{e}n wave packet.

The rest of the paper is organized as follows.  We briefly review the basic
equations of Magnetohydrodynamics. We first linearize the
basic ideal MHD equations to write down a wave equation for velocity
perturbation. Then, we recall some basic properties of its wave
solutions, i.e., MHD waves that includes sound and Alfv\'{e}n waves.
We employ the canonical quantization scheme
to introduce the concept of {\textit Alphenon} as the quantum of the
transverse and longitudinal Alfv\'{e}n wave, we study the thermodynamics of an
Alphonon system . Finally, we present our conclusion.
\section*{The basic equations of MHD}
\label{sec2}

Magnetohydrodynamics is a fluid theory that provides a theoretical framework to study the
macroscopic behavior of plasmas. In this framework, the behavior of a continuous plasma with mass
density ($\rho$), pressure ($p$), velocity $(\bfv)$, and current density $(\bfJ)$ is governed by
the set of coupled equations: ``\textit{a simplified form of Maxwell's equations, Ohm's Law,
	a gas law and equations of mass continuity, motion and energy}''~\citep{Priest}. If we neglect
the gravitational and all other dissipative forces from our
analysis, the ideal MHD equations in the non-relativistic limit are
given by~\citep{Priest,benz}
\begin{eqnarray}
&&\frac{\partial\rho}{\partial t}+\bf{\nabla}\cdot(\rho \bfv)=0, \label{e1}\\
&&\rho\frac{\partial \bfv}{\partial t}+\rho(\bfv\cdot\bf{\nabla})\bfv+\bf{\nabla} p=\frac{1}{\mu}\,(\bf{\nabla}\times\bfB)\times\bfB,  \label{e2}\\
&&\frac{\partial \bfB}{\partial t}=\bf{\nabla} \times(\bfv\times\bfB), \label{e3}\\
&&\bf{\nabla}\cdot\bfB=0, \label{e4}\\
&&p\rho^{-\gamma}={\rm const},\label{e5}
\end{eqnarray}
where $\bfB$, $\mu$, and $\gamma$ are the magnetic induction, magnetic permeability,
and the adiabatic exponent, respectively. Throughout the article, all vectors are denoted
as boldface. These coupled equations serve to determine $\rho$, $p$, $\bfv$ and $\bfB$ that can be
used to find other variables in plasma. For example, for the current density ($\bfJ$) and
the electric field ($\bfE$) we have, respectively, $\bfJ=\bf{\nabla}\times\bfB/\mu$
and $\bfE=-\bfv\times\bfB$.

\subsection*{MHD waves in a uniform unbounded magnetized plasma}
\label{sec3}

\begin{figure}[ht]
	\centering
	\includegraphics[scale=0.70]{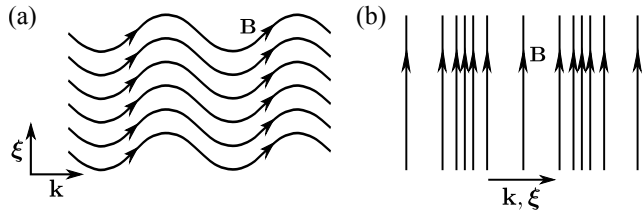}
	\caption{The sketch of the magnetic field lines, and the direction of the wave vector ($\bf{k}$), and the plasma eigendisplacement
		vector for (a) transverse Alfv\'{e}n wave, and (b) longitudinal Alfv\'{e}n wave.}
	\label{fig:1}
\end{figure}

We recall that for a uniformly magnetized medium the equilibrium
quantities such as mass density ($\rho_0$), pressure ($p_0$),
temperature ($T_0$), and magnetic field ($\bfB_0$) are constants.
The ideal MHD equations~(\ref{e1})--(\ref{e5}), can be linearized
about the equilibrium state. Indeed, the equilibrium quantities can
be perturbed as
\begin{eqnarray}
&& \bfB=\bfB_0+\bfB_1(\bfx, t), \label{e6}\\
&& p=p_0+p_1(\bfx, t), \label{e7}\\
&& \rho=\rho_0+\rho_1(\bfx, t), \label{e8}\\
&& \bfv=\bfv_1(\bfx, t),\label{e9}
\end{eqnarray}
where the perturbed mass density ($\rho_1$), pressure ($p_1$), magnetic field ($\bfB_1$),
and velocity ($\bfv_1$) are assumed to be small and the plasma initial flow is considered
in the static condition $(\bfv_0=0)$. Also, in our analysis below, the magnetic field is assumed
to be in the $z$-direction, i.e., $\bfB_0=B_0 \mathbf{z}$ (see Figure~\ref{fig:1}).

Inserting equations (\ref{e6})--(\ref{e9}) into
equations (\ref{e1})--(\ref{e5}), we get~\citep{Priest,benz}
\begin{eqnarray}
&& \frac{\partial\rho_1}{\partial t}+\rho _0(\bf{\nabla}\cdot\bfv_1)=0, \label{e10}\\
&& \rho_{0}\frac{\partial \bfv_1}{\partial t}+\bf{\nabla} p_1=
\frac{1}{\mu}(\bf{\nabla}\times\bfB_1)\times\bfB_0, \label{e11}\\
&&\frac{\partial \bfB_1}{\partial t}=\bf{\nabla} \times (\bfv_1 \times \bfB_0),  \label{e12}\\
&&\bf{\nabla}\cdot\bfB_1=0,\label{e14}\\
&&{p_1}={c}_{s}^{2}{\rho_1}, \label{e13}
\end{eqnarray}
where $c_s=\left(\frac{\gamma p_0}{\rho_0}\right)^{1/2}$ is the acoustic speed or sound speed.
These equations can be combined
to obtain the following wave equation for the disturbed velocity ($\bfv_1$)
\begin{equation}\label{e15}
\frac{\partial^2 \bfv_{1}}{\partial t^2}={c}_{s}^{2}\bf{\nabla} (\bf{\nabla}\cdot\bfv_1)+
\left\{\bf{\nabla}\times\left[\bf{\nabla} \times(\bfv_1 \times \bfB_0)\right]\right\}
\times \frac{\bfB_0}{\mu\rho_0}.
\end{equation}
which supports plane-wave solutions of the form
\begin{equation}\label{planewave}
\bfv_1(\bfx, t)=\hat{\bfv}_{1\bfk}\,e^{i(\bfk\cdot\bfx-\omega t)},
\end{equation}
where $\hat{\bfv}_{1\bfk}$, $\bfk$, and $\omega$ represent the Fourier amplitude of $\bfv_1$,
wave vector, and angular frequency, respectively.
Indeed, equation (\ref{e15}) together with equations (\ref{e10})--(\ref{e13}) provide a ground to study the
basic characteristics of the sound and MHD waves (Alfv\'{e}n, fast, and slow magnetoacoustic
waves) in a homogenous plasma that can be found in the literature
(see, for example, \citep{Priest,benz}). Here, to fix the notation, we briefly recall some basic properties of these main types of wave solution.

\subsection*{Sound wave}
\label{}

For a compressible plasma with no background magnetic field
$(\bfB_0=0)$, the only restoring force is the pressure gradient and
the wave equation (\ref{e15}) is reduced to the well-known sound wave
equation
\begin{equation}\label{e16}
\frac{\partial^2 \bfv_{1}}{\partial t^2}={c}_{s}^{2}\bf{\nabla} (\bf{\nabla}\cdot\bfv_1).
\end{equation}
Substituting the plane-wave solution equation (\ref{planewave}) in
equation (\ref{e16}) we find
\begin{equation}\label{e17}
\omega^2\hat{\bfv}_{1\bfk}=c_s^2\,\bfk(\bfk\cdot\hat{\bfv}_{1\bfk}),
\end{equation}
which, for $\bfk\cdot\bfv_{1\bfk}\neq 0$, yields the dispersion relation for acoustic waves or
sound waves, i.e., $\omega^2=c_s^2 k^2=k^2\left(\frac{\gamma p_0}{\rho_0}\right)$. Therefore, both
the phase speed ($v_{\mathrm ph}=\frac{\omega}{k}$) and group velocity
($v_g=\frac{\partial\omega}{\partial k}$) are equal to $\pm c_s$. Expectedly, as can be seen from
equation (\ref{e17}), sound waves are longitudinal (i.e., $\hat{\bfv}_{1\bfk} \| \bfk$).

The energy density (per unit volume) of a sound wave with frequency
$\omega$ is given by
\begin{equation}\label{e18}
\calU=\frac{1}{2}\rho_0 {\bfv}^{2}_{1}=\frac{1}{2}\rho_0 \omega^2\mbox{\boldmath $\xi$}^2,
\end{equation}
where, $\mbox{\boldmath $\xi$}$ is the eigendisplacement  of the wave equation
$\bfv_1=\frac{\partial \mbox{\boldmath $\xi$}}{\partial t}$.

\subsection*{Transverse Alfv\'en wave}

In an incompressible magnetized plasma ($\bf{\nabla}\cdot\bfv_1=0$), if
we make the assumption that the pressure remains uniform ($\bf{\nabla}
p=0$), we can restrict ourselves to the waves propagating along the
background magnetic field $(\mathbf{k}\parallel\bfB_0)$. If we take
$\bfB_1=\hat{\bfB}_{1\bfk} e^{i(\bfk\cdot\bfx-\omega t)}$,
equations (\ref{e11}) and (\ref{e12}) imply that the vectors $\bfv_1$ and
$\bfB_1$ are perpendicular to $\bfB_0$ (and also to $\bfk$).
Therefore, equation (\ref{e15}) can be reduced to
\begin{equation}\label{e19}
(\frac{k^2 {B}^{2}_{0}}{\mu}-\rho_0\omega^2){\hat{\bfv}_{1\bfk}}=0,
\end{equation}
which, for non-trivial solutions ($\hat{\bfv}_{1\bfk}\neq0$), gives
\begin{equation}\label{e20}
v^{2}_{ph}=\frac{\omega^2}{k^2}=\frac{{B}^{2}_{0}}{\mu\rho_0}\equiv{v}_{A}^{2},
\end{equation}
where ${v}_{ph}$ and ${v}_{A}$ are the phase speed and Alfv\'{e}n speed, respectively.

The dispersion relation (equation (\ref{e20})) and the propagation of the
wave along the magnetic field are two important characteristics  of
the transverse or shear Alfv\'{e}n wave. Using equations (\ref{e10}),
(\ref{e12}), and (\ref{e13}), we obtain
\begin{eqnarray}
&&\rho_1=0,\hspace{0.5cm}\hspace{0.5cm}p_1=0,  \label{e21}\\
&&\bfB_1=-\frac{1}{\omega}\bfk\times(\bfv_1\times\bfB_0)=
-\frac{\mathrm{sign}(k)k{B}_0}{\omega}\bfv_1,~~~\label{22}
\end{eqnarray}
where ``sign" is the \emph{sign} function. Equation~(\ref{e20}) shows that the shear Alfv\'{e}n wave is a
purely magnetic wave and, in view of equation (\ref{e11}), its restoring force is magnetic tension that
depends on the perturbed magnetic field~\citep{Priest,benz}.

The Alfv\'{e}n wave carries energy at the Alfv\'{e}n speed along the
magnetic field. The  average (over a period of oscillation) kinetic
energy density carried by the Alfv\'{e}n wave is
\begin{equation}\label{e23}
\calU_{\,\mathrm{kin.}}=\frac{1}{2}\rho_0 \langle\bfv_1^2\rangle
=\frac{1}{4}\rho_0\omega^2 {\mbox{\boldmath $\xi$}}^2.
\end{equation}
The time-averaged magnetic energy density related to the Alfv\'{e}n
wave is given by
\begin{equation}\label{e24}
\calU_{\,\mathrm{mag.}}=\frac{1}{2 \mu}\langle\bfB_1^2\rangle=
\frac{B_0^2}{2\mu}\frac{\langle\bfv_1^2\rangle}{v_A^2}=
\frac{B_0^2}{4\mu}\frac{\omega^2}{{v}^{2}_{A}}\,{\mbox{\boldmath $\xi$}}^2.
\end{equation}
Using the definition of the Alfv\'{e}n velocity  ($v_A$), we observe
that $\calU_{\,\mathrm{kin.}}=\calU_{\,\mathrm{mag.}}$. Therefore,
each mode of the transverse Alfv\'{e}n waves, with a given wave
vector $\bfk$ and frequency $\omega_\bfk$, may be treated as an
independent harmonic oscillator whose total average energy density
can be expressed by
\begin{equation}\label{e25}
\calU_{\,\bfk}=\calU_{\,\mathrm{kin.}} + \calU_{\,\mathrm{mag.}}=
\frac{1}{2}\rho_0\,\omega_\bfk^2\,{\mbox{\boldmath $\xi$}}^2.
\end{equation}
\subsection*{Longitudinal Alfv\'{e}n wave}\label{LAlfven}

In a compressible magnetized plasma $(\bf{\nabla}\cdot\bfv_1\neq0)$,
both the magnetic pressure and plasma pressure gradient forces
exist. These two restoring forces can act together to drive
longitudinal magnetoacoustic waves ($\bfk\parallel \bfv_1$) that
propagate perpendicular to the background magnetic field $\bfB_0$
(i.e., $\bfk\perp\bfB_0$) \citep{Priest,benz}. For longitudinal
plane wave solutions (\ref{planewave}), equation (\ref{e15}) can be
reduced to
\begin{equation}\label{e26}
\left(-\rho_0\omega^2+\gamma p_0k^2+\frac{k^2\bfB_0^2}{\mu}\right)
\,\hat{\bfv}_{1\bfk}=0,
\end{equation}
which implies
\begin{equation}\label{e27}
v_{\mathrm{ph}}^2=\frac{\omega^2}{k^2}=c_s^2+v_A^2.
\end{equation}
This dispersion relation represents a fast longitudinal MHD wave which propagates perpendicular
to the background magnetic field.

For small sound speed ($c_s\rightarrow 0$) the phase speed in equation (\ref{e27}) is equal to the
Alfv\'{e}n velocity, and then the corresponding wave is called the longitudinal or compressional
Alfv\'{e}n wave~\citep{benz}. In a similar manner, the same relation as given by equation (\ref{e25})
can be obtained for the average energy density carried by each mode of the longitudinal
Alfv\'{e}n waves.

\section*{Alphonons as the quanta of the Alfv\'{e}n waves}
\label{sec4}

\subsection*{Transverse Alphonons}
\label{subsec:rayleigh}

A classical dynamical theory of field with a set of well-defined
canonically conjugate variables can be quantized by promoting the
phase space variables to the Hermitian quantum operators. Then, the
field and the Hamiltonian of the system can be expressed in terms of
the creation and annihilation operators that defines the excitations
(quanta) of the filed. The tensor product of the one-particle states
make a basis for the Hilbert space (Fock space) of the quantum
theory.

Because the particles in transverse Alfv\'{e}n waves oscillate in a
plane perpendicular to the background magnetic field (and also to
the direction of propagation), each mode of oscillation with a given
wave vector $\bfk$ may be described by two distinct polarization
vectors $\mbox{\boldmath $\varepsilon$}^\lambda_\bfk$ with $\lambda=1, 2$. This is
identical to the physical polarizations of electromagnetic
fields \citep{Schwartz,Ryder}. If we take $z$-direction as the
direction of propagation, i.e., $\bfk=k\,\bf{z}$, then two
transverse linear polarization vectors can be read as
\begin{eqnarray}\label{e29}
\mbox{\boldmath $\varepsilon$}^1_\bfk=(1, 0, 0),~~~~~~~~~\mbox{\boldmath $\varepsilon$}^2_\bfk=(0, 1, 0).
\end{eqnarray}

Consequently, for each Alfv\'{e}n mode with a given $\bfk$, the plasma displacement vector $\mbox{\boldmath $\xi$}$
and hence the disturbed velocity $\bfv_1$ may be expressed, respectively, by
$\mbox{\boldmath $\xi$}=\xi_{\bfk, 1}\,\mathbf{\varepsilon}^1_\bfk+\xi_{\bfk, 2}\,\mathbf{\varepsilon}^2_\bfk$ and
$\bfv_1=v_{\bfk, 1}\,\mbox{\boldmath $\varepsilon$}^1_\bfk+v_{\bfk, 2}\,\mbox{\boldmath $\varepsilon$}^2_\bfk.$

Next, we observe that equation (\ref{e25}) can be rewritten as
\begin{equation}\label{e33}
\calU_{\,\bfk}=\frac{1}{2}\left(P_{\bfk, \lambda}^2+\omega_\bfk^2 Q_{\bfk, \lambda}^2\right),
\end{equation}
provided that we introduce a pair of canonical variables, $Q_{\bfk,
	\lambda}$ and its conjugate momentum $P_{\bfk, \lambda}$, such that
\begin{eqnarray}
&&Q_{\bfk, \lambda}=\sqrt{\rho_0}\left(\xi_{\bfk, \lambda}+\xi_{\bfk, \lambda}^*\right),
\label{e31}\\
&&P_{\bfk, \lambda}=\frac{d Q_{\bfk, \lambda}}{dt}=
-i\omega_\bfk\sqrt{\rho_0}\left(\xi_{\bfk, \lambda}-\xi_{\bfk, \lambda}^*\right),\label{e32}
\end{eqnarray}
where ``$^*$" stands for the complex conjugation. Equation~(\ref{e33}) represents the energy of a
harmonic oscillator with unit mass, position variable $Q_{\bfk, \lambda}$ and momentum $P_{\bfk, \lambda}$.

To quantize the velocity field (or equivalently the displacement
vector), we promote the canonical variables $Q_{\bfk, \lambda}$ and
$P_{\bfk, \lambda}$ to Hermitian operators and introduce the
annihilation operator
\begin{equation}\label{e34}
a_{\bfk, \lambda}=\frac{1}{\sqrt{2\hbar\,\omega_\bfk}}\left(\omega_\bfk
Q_{\bfk, \lambda} + i P_{\bfk, \lambda}\right),
\end{equation}
for the plane wave modes with the wave vector $\bfk$ and frequency
$\omega_\bfk$. The adjoint of equation (\ref{e34}) defines the creation
operator $a^\dagger_{\bfk, \lambda}$. The creation and annihilation
operators $a_{\bfk, \lambda}$ and $a^\dagger_{\bfk, \lambda}$
satisfy the (bosonic fields) communication relations
\begin{eqnarray}\label{e35}
&&[a_{\bfk, \lambda},\, a^\dagger_{\bfk',\lambda'}]=
\delta_{\lambda,\lambda'}\,\delta(\bfk-\bfk'), \\\nonumber
&&\left[a_{\bfk, \lambda},\, a_{\bfk', \lambda'}\right]=
[a^\dagger_{\bfk, \lambda},\, a^\dagger_{\bfk', \lambda'}]=0, 
\end{eqnarray}
where $\delta(\bfk-\bfk')$ denotes the  Dirac delta function. In
terms of these operators, the energy density (\ref{e33}) can be
expressed by the Hamiltonian
\begin{equation}\label{e36}
\hat{\calH}_{\,\bfk}=\sum_{\lambda=1}^{2}\hbar\,\omega_\bfk\left(a^\dagger_{\bfk, \lambda}
a_{\bfk, \lambda} + \frac{1}{2}\right).
\end{equation}
This suggests that the state with $n_{\bfk, \lambda}$ quanta of
Alfv\'{e}n wave in ($\bfk, \lambda$)-mode can be read as
\begin{equation}\label{e37}
|\bfk, \lambda, n_{\bfk, \lambda}\rangle=\frac{1}{\sqrt{n_{\bfk, \lambda}!}}
\left(a_{\bfk, \lambda}^\dagger\right)^{n_{\bfk, \lambda}}
|\bfk, 0\rangle,
\end{equation}
where $|\bfk, 0\rangle$ is the vacuum  state for a given mode
$\bfk$. These are also the eigenstates of the \textit{Alphonon}
occupation number density operator defined by
$\hat{\mathcal{N}}_{\bfk, \lambda}=a^\dagger_{\bfk, \lambda}
a_{\bfk, \lambda}$. We then note that the expectation value of
$\hat{\mathcal{N}}_{\bfk, \lambda}$ given by
\begin{equation}\label{e38}
\mathcal{N}_{\bfk}=\sum_{\lambda=1}^2\langle a^\dagger_{\bfk, \lambda} a_{\bfk, \lambda}\rangle =
\frac{\rho_0}{2\hbar}\,\omega_\bfk{|\mbox{\boldmath $\xi$}_\bfk|}^2=
\frac{\rho_0}{2\hbar}\,{|\hat{\bfv}_{1\bfk}|}^2,
\end{equation}
that gives Alphonon occupation number, can be used to characterize
the Alphonons concentration density. Equation(\ref{e38}) shows that a
single Alphonon corresponds to a plasma fluctuation $\mbox{\boldmath $\xi$}_\bfk$ (or
equivalently, velocity fluctuation $\bfv_\bfk$), with an amplitude
given by
\begin{equation}\label{e39}
{|\mbox{\boldmath $\xi$}_\bfk|}_{\mathrm{min}}=\left(\frac{2\hbar}{\rho_0\omega_\bfk}\right)^{\!\frac{1}{2}}
~~~ \Longleftrightarrow ~~~
{|\hat{\bfv}_{1\bfk|}}_{\mathrm{min}}=\left(\frac{2\hbar\,\omega_\bfk}{\rho_0}\right)^{\!\frac{1}{2}},
\end{equation}
where $\omega_\bfk=k\,v_A=k B_0/\sqrt{\mu \rho_0}$ is the Alfv\'{e}n wave frequency.
This establishes the characteristic scale of the displacement (or velocity) quantization in a
magnetized plasma.

Finally,  we note that a general transverse Alfv\'{e} wave packet can be obtained by
superimposing many plane-wave solutions. In view of equations (\ref{e31}) and (\ref{e32}),
we observe that
\begin{equation}
\xi_{\bfk, \lambda}=\left(\frac{\hbar}{2\rho_0\omega_\bfk}\right)^{\!\frac{1}{2}}\!\!
a_{\bfk, \lambda},~~~~~
\xi_{\bfk, \lambda}^*=\left(\frac{\hbar}{2\rho_0\omega_\bfk}\right)^{\!\frac{1}{2}}\!\!
a^\dagger_{\bfk, \lambda}.
\label{ez1}
\end{equation}
Thus, we can express a general displacement vector $\bf{\xi}(\bfx, t)$, describing a general Alfv\'{e}n
wave packet, according to (Note that for transverse Alfv\'{e}n mode we take
$\bfk=k\mathbf{z}$ in the direction of background magnetic field $\bfB_0$.)
\begin{equation}\label{ez5}
\mbox{\boldmath $\xi$}(\bfx, t)=\int\!\!\frac{d\,k}{\sqrt{2\pi}}\!
\left(\frac{\hbar}{2\rho_0\omega_\bfk}\right)^{\!\frac{1}{2}}\!
\sum_{\lambda=1}^{2} {\mbox{\boldmath $\varepsilon$}}^\lambda_\bfk\,a_{\bfk, \lambda}\,
e^{i(\bfk\cdot\bfx-i\omega_\bfk t)},
\end{equation}
which yields the velocity fluctuation
\begin{equation}\label{ez55}
\bfv_1(\bfx, t)=-i\!\int\!\!\frac{d\,k}{\sqrt{2\pi}}\!
\left(\frac{\hbar \omega_\bfk}{2\rho_0}\right)^{\!\frac{1}{2}}\!
\sum_{\lambda=1}^{2} {\mbox{\boldmath $\varepsilon$}}^\lambda_\bfk\,a_{\bfk, \lambda}\,
e^{i(\bfk.\bfx-i\omega_\bfk t)}.
\end{equation}
For this wave packet we may introduce a quantum field $\bfQ(\bfx, t)$ and its conjugate momentum $\bfP(\bfx, t)$
according to
\begin{eqnarray}
&&\bfQ(\bfx, t)=\left(\frac{\hbar\rho_0}{2}\right)^{\!\frac{1}{2}}\!\!\!\int\!\!\!
\frac{d\,k}{\sqrt{2\pi}}\,\frac{1}{\sqrt{\omega_\bfk}}\sum_{\lambda=1}^{2}\!
\left(\mbox{\boldmath $\varepsilon$}^\lambda_\bfk\,a_{\bfk, \lambda}\,e^{ik\cdot x}+\,{\mbox{\boldmath $\varepsilon$}}^{\lambda *}_{\bfk}\,a^\dagger_{\bfk, \lambda}
\,e^{-ik\cdot x}\right)\!,\label{ezz}\\
&&\bfP(\bfx, t)=-i\left(\frac{\hbar\rho_0}{2}\right)^{\!\frac{1}{2}}\!\!\!\int\!\!\!
\frac{d\,k}{\sqrt{2\pi}}\,\sqrt{\omega_\bfk}\sum_{\lambda=1}^{2}
\left(\mbox{\boldmath $\varepsilon$}^\lambda_\bfk\,a_{\bfk, \lambda}\,e^{ik\cdot x}-\,{\mbox{\boldmath $\varepsilon$}}^{\lambda *}_{\bfk}\,a^\dagger_{\bfk, \lambda}
\,e^{-ik\cdot x}\right)\!,
\end{eqnarray}
where, $k\cdot x=\bfk\cdot\bfx-\omega_\bfk t$. Making use of
equation (\ref{e35}), we can show that these fields satisfy canonical
commutation relation, i.e.,
\begin{equation}\label{PQcom}
\left[\bfQ_i(\bfx, t),\, \bfP_j(\bfx', t)\right]= i\hbar\,\delta_{ij}\,\delta(\bfx-\bfx').
\end{equation}

\subsection*{Longitudinal Alphonons}
\label{}

As discussed in Subsec.~\ref{LAlfven}, for the longitudinal Alfv\'{e}n wave, the direction of plasma
oscillations is parallel to the direction of wave propagation (the wave vector $\bfk$). Therefore,
a single polarization vector $\mathbf{\varepsilon}_\bfk$ along the wave vector $\bfk$ is required to express
the oscillation amplitude. More explicitly, we take $\mbox{\boldmath $\varepsilon$}_\bfk=\hat{\bfk}=\bfk/k$ and write
$\mbox{\boldmath $\xi$}=\xi_\bfk \mbox{\boldmath $\varepsilon$}_\bfk$, and $\bfv_1=v_{\bfk,1} \mbox{\boldmath $\varepsilon$}_\bfk$.

To quantize the longitudinal Alfv\'{e}n wave or its corresponding
velocity field $\bfv_1$, we define a pair of canonical variables,
$Q_\bfk$ and  $P_\bfk$, by
\begin{eqnarray}
&&Q_\bfk=\sqrt{\rho_0}\left(\xi_\bfk+\xi_\bfk^*\right),\label{e40}\\
&&P_\bfk=\frac{d Q_\bfk}{d t}=-i\omega_\bfk\sqrt{\rho_0}
\left(\xi_\bfk-\xi_\bfk^*\right)\label{e41}.
\end{eqnarray}
Making use of these variables into equation (\ref{e25}), we find
$\calU_{\,\bfk}=\frac{1}{2}\left(P_\bfk^2+\omega_\bfk^2
Q_\bfk^2\right)$ that describes a harmonic oscillator with unit mass
and frequency $\omega_\bfk$. Now, it is convenient to introduce the
annihilation operator
\begin{equation}\label{e43}
a_\bfk=\frac{1}{\sqrt{2\hbar\,\omega_\bfk}}\left(\omega_\bfk Q_\bfk + i P_\bfk\right),
\end{equation}
that together with its adjoint satisfy the commutation relations
\begin{equation}\label{e44}
[a_\bfk,\, a^\dagger_{\bfk'}]=\delta(\bfk-\bfk'),~~~~~
~~~~~[a_\bfk,\, a_{\bfk'}]=[a^\dagger_\bfk,\,a^\dagger_{\bfk'}]=0.
\end{equation}
These simplify the Hamiltonian of the $\bfk$-mode longitudinal
Alfv\'{e}n wave to the form
\begin{equation}\label{e45}
\calH_\bfk=\hbar\,\omega_\bfk\left(a^\dagger_\bfk a_\bfk+\frac{1}{2}\right),
\end{equation}
which has the general eigenvector $|\bfk,n_\bfk\rangle$ represented
by
\begin{equation}\label{e46}
|\bfk,n_\bfk\rangle=\frac{1}{\sqrt{n_\bfk!}}\left(a^\dagger_\bfk\right)^{n_\bfk}|\bfk,0\rangle.
\end{equation}
These states occupy $n_\bfk$ quanta (Alphonons) of the longitudinal Alfv\'{e}n wave with frequency
$\omega_\bfk$ and longitudinal polarization ($\mathbf{\varepsilon}_\bfk||\bfk$). Accordingly, the longitudinal
Alphonon occupation number and the corresponding characteristic velocity fluctuation can be read,
respectively, by the same expressions as given in equations (\ref{e38}) and (\ref{e39}).

\section*{Thermodynamics of an Alphonon system}
\label{sec6}

We assume that the system of Alphonons is in thermal contact with the environment at temperature
$T$. Also, we suppose both the volume of the system and the number of particles (number density)
are constant. A collection of such systems with several copies is called the canonical ensemble.
The canonical partition function depends on the degree of freedom of the system in both classical
and quantum mechanics \citep{Landau,pathriastatistical,huang1963statistical}.
For the canonical ensemble, the partition function is given by
\begin{eqnarray}\label{e51}
&&Z=\sum_{\bfk}\exp\left(-\frac{E_{\bfk}}{k_{B}T}\right)\\\nonumber
&&\hspace{0.35cm}={\bf \sum_{n_1=0}^{\infty}\cdots\sum_{n_{{\bfk}_{max}}=0}^{\infty}\exp\left(-\frac{1}{k_BT}\sum_{{\bfk}=1}^{{\bfk}_{max}}
	\hbar\omega_{\bfk}\left(n_{\bfk}+\frac{1}{2}\right)\right) }\\\nonumber
&&\hspace{0.35cm}=\prod_{\bfk=1}^{\bfk_{\max}}\frac{\exp\left(\frac{-\hbar\omega_{\bfk}}{2k_{B}T}
	\right)}{1-\exp\left(\frac{-\hbar\omega_{\bfk}}{k_{B}T}\right)}.
\end{eqnarray}
where $\hbar$, $k_{B}$, $T$, and ${\bfk}_{max}$ are, respectively, the Planck's constant,
Boltzmann constant, temperature, and maximum wave number of the Alphonon in mode $\bfk$.

Using equation (\ref{e51}), the Helmhholtz free energy $F$ is defined by
\begin{eqnarray}\label{e52}
&&F=-{k}_{B}T\ln Z\\\nonumber
&&\hspace{0.35cm}=\sum_{\bfk}\frac{\hbar\omega_{\bfk}}{2}+k_{B}T\sum_{\bfk}\ln
\left(1-\exp\left(\frac{-\hbar\omega_{\bfk}}{k_{B}T}\right)\right).
\end{eqnarray}
The first term in the right hand side is the zero-temperature energy of the system. For more
simplicity we remove the zero-temperature from our analysis. Moreover, the specific energy per
volume is introduced by
\begin{eqnarray}\label{e54}
&&f=\frac{1}{L^3}F=\frac{{k}_{B}T}{L^3}\sum_{\bfk}\ln \left(1-\exp\left(\frac{-\hbar\omega_{\bfk}}{k_{B}T}\right)\right)\\\nonumber
&&\hspace{0.30cm}\approx {k}_{B}T \int_{0}^{\infty}D(\omega)\ln \left(1-\exp(\frac{-\hbar\omega}{{k}_{B}T})\right) d\omega,
\end{eqnarray}
where the density of states in frequency $D(\omega)$ is related to the mode occupies the volume
$\frac{2\pi}{L^3}$ in the wave vector space by
\begin{equation}\label{e55}
D(\omega)=D(k)\frac{dk}{d \omega},\hspace{1.5cm}D(k)dk=\frac{k^2}{2\pi^2}dk.
\end{equation}
Using the dispersion relation $\frac{dk}{d\omega}$, the dencity of state can be read as
\begin{equation}\label{e57}
D(\omega)=\frac{\omega^2}{2\pi^2}{\left(\frac{\rho_0\mu}{{B}^{2}_{0}}\right)}^{\frac{3}{2}}.
\end{equation}

For the entropy per volume $s$ of the Alphonon system we have
\begin{eqnarray}\label{e58}
&&s=-\frac{\partial f}{\partial T}=-k_B\int_{0}^{\infty}\frac{\omega^2}{2\pi^2}{
	\left(\frac{\rho_0\mu}{{B}^{2}_{0}}\right)}^{\frac{3}{2}}\left(\ln \left(1-\exp(\frac{-\hbar\omega_\bfk}{k_{B}T})\right)+
\frac{\hbar\omega\exp(\frac{-\hbar\omega}{k_{B}T})}{k_BT(\exp(\frac{-\hbar\omega}{k_{B}T})-1)}
\right) d\omega.
\end{eqnarray}
Also, the internal energy $u$ is defined by
\begin{equation}\label{e59}
u=f+T.s=\int_{0}^{\infty}\frac{\omega^2}{2\pi^2}{\left(\frac{\rho_0\mu}{{B}^{2}_{0}}\right)}^{\frac{3}{2}}\frac{\hbar\omega}{(\exp(\frac{\hbar\omega}{k_{B}T})-1)}d\omega.
\end{equation}
According to the expression $u=\int_{0}^{\infty}u(\omega)d \omega$, equation (\ref{e59}) gives the
spectral energy density as
\begin{equation}\label{e60}
u(\omega)=\frac{\omega^2}{2\pi^2}{\left(\frac{\rho_0\mu}{{B}^{2}_{0}}\right)}^{\frac{3}{2}}\frac{\hbar\omega}{\exp\left(\frac{\hbar\omega}{k_{B}T}\right)-1}.
\end{equation}
Equation (\ref{e60}) is equivalent to the  Planck's black body radiation law for an Alfv\'{e}n wave system.

Finally, for the total internal energy $u$ we obtain
\begin{equation}\label{e65}
u={\left(\frac{\rho_0\mu}{{B}^{2}_{0}}\right)}^\frac{3}{2}\frac{\pi^2{k}^{4}_{B}}{30\hbar^3}T^4.
\end{equation}
Then, using equation (\ref{e65}), the specific heat per volume is given by
\begin{equation}\label{e64}
c_V=\frac{\partial u}{\partial T}={\left(\frac{\rho_0\mu}{{B}^{2}_{0}}\right)}^\frac{3}{2}
\frac{2\pi^2{k}^{4}_{B}}{15\hbar^3}T^3.
\end{equation}

\section*{Conclusion}
\label{sec5}

In a gaseous plasma or a solid-state plasma, the propagation of the
Alfv\'{e}n waves is a significant characterization of the disturbed
magnetic field. The Alfv\'{e}n waves has been detected in astrophysical,
laboratory, and solid-state plasmas. The magnetic pressure and
tension are the restoring forces of the longitudinal and transverse
Alfv\'{e}n waves, respectively. Both the transverse and longitudinal
Alfv\'{e}n waves carry an average energy density that is, for each
mode of oscillation, proportional to the square of the velocity
perturbation (and hence to the square of the plasma  displacement
vector) and is reminiscent of the classical energy of a harmonic
oscillator with unit mass. This observation suggests that each mode
of the Alfv\'{e}n wave may be treated as an independent harmonic
oscillator that can be canonically quantized. In particular, we
introduce a pair of canonical variables that specify the phase space
of the underlying classical system and use the Dirac's canonical
quantization prescription to write down the Alfv\'{e}n wave velocity
filed (or the corresponding plasma displacement vector) in terms of
the creation and annihilation operators. This defines the concept of
``{\emph Alphonons}", the elementary excitations of the Alfv\'{e}n
wave velocity in a magnetized plasma that was first introduced in~\cite{gheibi}.

We studied the characteristic features of the Alfv\'{e}n wave quanta, the Alphonons. Specifically,
we showed that two polarization vectors (perpendicular to the wave vector) are required for the
quantization of the transverse Alfv\'{e}n wave. These two physical polarization vectors and the
corresponding dispersion relation (see equation (\ref{e20})) closely resemble those of
the photon. This fact allows us to assign spin one to the Alphonon as a quantum of the transverse
Alfv\'{e}n wave. We also used a single polarization vector (parallel to the wave vector) to quantize
the longitudinal Alfv\'{e}n wave. This means that the longitudinal Alfv\'{e}n wave quanta can be
specified with a quantized scalar filed (the amplitude of the velocity field or the displacement
vector along the wave vector). Therefore, these quanta have zero spin. This shows that the
Alphonons obey the bosons behavior. Indeed, any number of identical excitations (normal modes of
the Alfv\'{e}n waves related to the perturbations of the magnetic fields and the reason for the
oscillatory motions for particles in plasma) can be created by the repeated application of the
creation operator  $a^\dagger$. For a given state of the system the creation operator increases
the number of particles by one in that state. This is equivalent to the excitation of new
Alfv\'{e}n mode in the system (new Alphonon).

Finally, we also studied the thermodynamic properties of a system of Alphonons. Assuming
that the system with a constant volume and fixed number of quasi-particles is in a thermal contact
with the environment at temperature $T$, we can consider the canonical ensemble and write down the
expressions for the Helmhholtz free energy. From the dispersion relation we determined the density
of states, the entropy, the total internal energy. Then the spectral energy density, and the
specific heat in constant volume are obtained. We showed that the  spectral energy density of the
Alphonon system coincides with the Planck's black body radiation law.

An immediate application of the concept of Alphonon was made by~\cite{gheibi}, who showed that in
the horizon of an Alfv\'{e}nic black hole, Alfv\'{e}n waves will be trapped and emit Hawking
radiation that comprises the Alphonons. Detecting the Hawking radiation from the gravitational
black holes is an important task for scientists. Because the temperatures predicted by Hawking
effect is far less than the cosmic microwave background radiation temperature,
it makes difficult to verify the black hole evaporation. Nevertheless, the concept of the artificial black holes (e.g.,
sonic, Alfv\'{e}nics, magnetoacoustic) is recently investigated to study the properties of Hawking
radiations. On the other hand, the solid-state plasma, solar and stellar atmospheres, the space plasma (e.g.,
intergalactic, interplanetary, interstellar medium, etc) are examples that the concept of Alphonons
may be useful to study the collective behavior of such systems.


\listofchanges
\end{document}